\documentclass[12pt]{article}

\usepackage{graphicx}
\usepackage{psfig}

\usepackage{scicite}

\usepackage{times}

\newenvironment{sciabstract}{%
\begin{quote} \bf}
{\end{quote}}

\newcounter{lastnote}
\newenvironment{scilastnote}{%
\setcounter{lastnote}{\value{enumiv}}%
\addtocounter{lastnote}{+1}%
\begin{list}%
{\arabic{lastnote}.}
{\setlength{\leftmargin}{.22in}}
{\setlength{\labelsep}{.5em}}}
{\end{list}}

\title{Switched magnetospheric regulation of pulsar spin-down}

\author
{Andrew Lyne,$^{1\ast}$ George Hobbs,$^{2}$ Michael Kramer,$^{1,3}$ Ingrid Stairs,$^{4}$ Ben Stappers$^{1}$\\
\\
\normalsize{$^{1}$ Jodrell Bank Centre for Astrophysics, School of
Physics and Astronomy,}\\
\normalsize{University of Manchester, Manchester, M13~9PL, UK}\\
\normalsize{$^{2}$ Australia Telescope National Facility, CSIRO,
PO~Box~76, Epping NSW~1710, Australia}\\
\normalsize{$^{3}$ MPI f\"ur Radioastronomie, Auf dem H\"ugel 69,
53121 Bonn, Germany}\\
\normalsize{$^{4}$ Dept. of Physics and Astronomy, University of British Columbia, 
6224 Agricultural Road,}\\
\normalsize{Vancouver, BC V6T~1Z1, Canada}\\
\normalsize{$^\ast$To whom correspondence should be addressed; E-mail:  andrew.lyne@manchester.ac.uk.}
}

\date{}

\begin{document} 


\baselineskip24pt


\maketitle


\begin{sciabstract}
Pulsars are famed for their rotational clock-like stability and their
highly-repeatable pulse shapes.  However, it has long been known that
there are unexplained deviations (often termed "timing noise") from
the rate at which we predict these clocks should run. We show that
timing behaviour often results from typically two
different spin-down rates.  Pulsars switch abruptly between these
states, often quasi-periodically, leading to the observed spin-down
patterns.  We show that for six pulsars the timing noise is correlated
with changes in the pulse shape.  Many pulsar phenomena including
mode-changing, nulling, intermittency, pulse shape variability and
timing noise are therefore linked and caused by changes in the
pulsar's magnetosphere.  We consider the possibility that
high-precision monitoring of pulse profiles could lead to the
formation of highly-stable pulsar clocks.
\end{sciabstract}


\section*{Introduction}

Neutron stars form in the supernova collapse of the cores of exhausted
massive stars and are comprised of some of the densest and most
extreme matter in the observable Universe.  Pulsars are
rapidly-rotating, highly-magnetized neutron stars.  As they rotate,
intense beams of electromagnetic radiation may sweep across the
Earth, resulting in pulses which are often observed with
radio telescopes, enabling their rotation to be studied with high
precision.

Pulsars are amongst the most stable rotators known in the Universe.
Over long time spans the fastest spinning pulsars known as
``millisecond pulsars'' even rival the stability of atomic
clocks\cite{pt96}.  Although they slow down gradually because of the
conversion of rotational energy into highly energetic particles and
electromagnetic waves, a simple spin-down model using only the
pulsar's rotational frequency $\nu$ and its first time derivative
$\dot{\nu}$ is often sufficient to reveal timing properties that, for
instance, allow high-precision tests of the theory of general
relativity \cite{ksm+06} and may also allow direct detection
of gravitational waves \cite{saz78,det79,jhlm05}.  However, not all
pulsars seem to be perfectly stable clocks.

The pulsar timing technique\cite{lk05,ehm06} is used to compare pulse
arrival times at an observatory with times predicted from a spin-down
model. Many pulsars show a phenomenon known as ``timing noise'' where
seemingly quasi-random walks in the rotational parameters are
observed. The largest study of such kind \cite{hlk10} recently
presented the rotation properties of 366 pulsars, measured mainly
using the Lovell Telescope at Jodrell Bank. This long-term monitoring
of pulsars over 40 years made it possible to study phenomena in many
pulsars over decadal timescales. It was shown that the majority of the
pulsars were found to have significant irregularities in their
rotation rate. The differences between the observed and predicted
times, known as the pulsar ``timing residuals'', can be less than a
few milliseconds over more than $30$\,yr, but in other cases timing
residuals can be as large as many seconds.  In contrast to the
standard models held for the past $\sim$40\,yr, it was found that
these timing irregularities are often quasi-periodic with long
($\sim$1-10 year) time scales. Here we present a description of these
irregularities and how they are related to changes in pulse shape,
linking many peculiar and unexplained time-dependent phenomena
observed in pulsars.

\section*{Pulsar Time Scales}

Many of the properties of pulsars are not perfectly stable and they
vary over a wide range of timescales.  Rotational periods range from
milliseconds to seconds.  The structure and brightness of individual
pulses are observed to vary significantly, but the average of many
hundreds of individual pulses ($\sim$minutes) is usually stable,
leading to a characteristic profile that is often unique to a
pulsar. On time scales of seconds to hours, some pulsars are observed
to exhibit either ``nulling'' events, during which the pulse emission
switches ``off'', or ``mode changing'' events where the observed pulse
profile changes abruptly between two (sometimes three) well-defined
shapes.  On longer time scales, PSR~B1931+24 has recently been
described as an ``intermittent pulsar'' which relates to the fact that
it undergoes extreme nulling events\cite{klo+06}, displaying
quasi-periodic behaviour in which the pulsar acts as a normal pulsar
for typically five to ten days, before switching ``off'', being
undetectable for $\sim$25 days and then abruptly switching ``on''
again.  On even longer time scales, stable harmonically-related
periodicities of $\sim$250, 500 and 1000\,d, have been reported in the
rotation rate and pulse shape of PSR~B1828$-$11. The periodicities
have generally been interpreted as being caused by PSR~B1828$-$11
freely-precessing\cite{sls00}, even though it had been argued that
this was not possible in the presence of the superfluid component
believed to exist inside neutron stars\cite{sha77}. 

\section*{Discrete pulsar spin-down states}

The patterns observed in the timing residuals of a sample of 10
pulsars (Fig.~1) are typical of the sample presented in \cite{hlk10}
and highlight the main results of that paper that 1) the residuals are
dominated by quasi-periodic structures and 2) the residuals are
generally asymmetric, in that the radii of curvature of local maxima
are often consistently different to those of local minima.  Clear
examples are seen in PSRs B0950+08, B1642$-$03, B1818$-$04, B1826$-$17
and B1828$-$11.  In several cases, notably PSRs B0919+06 and B1929+20,
relatively rapid oscillations lie on lower-frequency structures.

Structures in the timing residuals have been widely discussed in the
literature.  Sudden increases in the pulsar's rotation rate are known
as ``glitches'' and are explained by the sudden unpinning of
superfluid vortices in the interior of the neutron star \cite{ai75}.
An apparently related phenomenon known as ``slow glitches'' has been
described\cite{zww+04,sha07}, characterised by a slow permanent
increase in rotation rate but no substantial change in the slow-down
rate and also identified with the interior of the neutron star.  The
low-frequency structures seen over short data spans were thought to
represent either random walks in the pulse frequency and/or its
derivatives \cite{bgh+72,cd85} arising from instabilities within the
neutron star superfluid interior, multiple micro-glitches \cite{js06},
free precession of the neutron star \cite{sls00}, asteroid belts
\cite{cs08}, magnetospheric effects\cite{che87}, interstellar or
interplanetary medium effects\cite{sfal97} or accretion of material
onto the pulsar's surface\cite{qxx+03}.  Timing residuals for the
youngest pulsars in \cite{hlk10} are dominated by the recovery from
glitch events, sometimes having occurred prior to the start of
observing.  In general, for the remaining pulsars it was shown that,
with long data spans, the low-frequency structures are no longer
dominant, but are now understood as restricted pieces of much
longer-term oscillatory structures, often with asymmetric maxima and
minima. Any model explaining timing noise therefore needs to explain
these commonly-occurring features.

The analysis of PSR~B1931$+$24\cite{klo+06} showed that the pulsar
spin-down rate switched by $\sim$50\% between the ``on'' and ``off''
states, with the pulsar spinning down faster when the radio signal was
detectable.  The quasi-periodic nature of the time between state
changes and the difference in time spent in each state leads naturally
to oscillatory, asymmetric timing residuals (Fig.~6).  The
existence of two discrete spin-down rates in PSR B1931$+$24 and the
similarities between such timing residuals and those shown in
Fig.~1 suggests that a similar model could be also
applied to the timing noise seen in all pulsars.

The variation in the spin-down rate
$\dot{\nu}(t)$ for 17 pulsars demonstrates that the
observed structures in the timing residuals arise from a pulsar's
$\dot{\nu}$ moving between a small number of values and frequently in
an oscillatory manner (Fig.~2).  In some cases more complex structure is seen.
For instance, in PSR~B1642$-$03 we observe peaks in $\dot{\nu}$
followed by a sudden change to a more negative $\dot{\nu}$ value
before a slow gradual rise. In PSR~B1828$-$11, in addition to the
oscillatory structure, we also observe a long-term gradual linear
change in $\dot{\nu}$ across the data span.  We
concentrate on the dominant features of this figure: the value of
$\dot{\nu}$ changes between a few (typically two) well-defined values,
often in a quasi-periodic manner.

In order to quantify the behaviour, we measured the peak-to-peak
values of $\dot{\nu}$ for each pulsar (Table~1).  Additionally, for
each of the time sequences in Fig.~2, we have performed
Lomb-Scargle\cite{sca82} and wavelet\cite{fos96} spectral analyses.
As expected, some of the resultant spectra (Figs.~7 and 8) show
narrow, highly-periodic features, while others show broader, less
well-defined peaks.

\begin{table*}[hbt]
\begin{center}
\caption{Measured parameters of 17 pulsars presented in Fig.~2, as
well as PSR~B1931+24 which is also discussed in the text. We give the
pulsar names, rotational frequency $\nu$ and the first derivative
$\dot{\nu}$, followed by the peak-to-peak fractional amplitude $\Delta
\dot{\nu}/\dot{\nu}$ of the variation seen in Fig.~2.  The pulsars are
given in order of decreasing value of this quantity.  We also present
the fluctuation frequencies F of the peaks of the Lomb-Scargle power
spectra (Fig.~7), with the widths of the peaks or group of peaks
given in parenthesis in units of the last quoted digit.}
\begin{tabular}{llrrrll}\\
\hline
\hline
\multicolumn{1}{c}{Pulsar} & \multicolumn{1}{c}{Jname} & \multicolumn{1}{c}{$\nu$} 
& \multicolumn{1}{c}{$\dot{\nu}$} & $\Delta \dot{\nu}/\dot{\nu}$ &
\multicolumn{1}{c}{F} & {Comment}\\
\multicolumn{1}{c}{name} & & (Hz) & (Hz s$^{-15}$) & (\%) & (yr$^{-1}$) & \\
\hline\\
B1931+24$^{a}$&J1933+2421 & 1.229 &  $-$12.25 &  44.90 &    13.1(7) & Intermittent pulsar \\
B2035+36   & J2037+3621   & 1.616 &  $-$12.05 &  13.28 &    0.02(2)  &   28\% change in W$_{eq}$ \\
B1903+07   & J1905+0709   & 1.543 &  $-$11.76 &   6.80 &    0.36(13) &    \\
J2043+2740 & J2043+2740   & 10.40 & $-$135.36 &   5.91 &    0.11(5)  &   100\% change in W$_{50}$ \\
B1822$-$09 & J1825$-$0935 & 1.300 &  $-$88.31 &   3.28 &    0.40(7)  &   100\% change in A$_{pc}$/A$_{mp}$ \\
B1642$-$03 & J1645$-$0317 & 2.579 &  $-$11.84 &   2.53 &    0.26(7)  & \\
B1839+09   & J1841+0912   & 2.622 &   $-$7.50 &   2.00 &    1.00(15) & \\
B1540$-$06 & J1543$-$0620 & 1.410 &   $-$1.75 &   1.71 &    0.24(2)  &   12\% change in W$_{10}$ \\
B2148+63   & J2149+6329   & 2.631 &   $-$1.18 &   1.69 &    0.33(7)  & \\
B1818$-$04 & J1820$-$0427 & 1.672 &  $-$17.70 &   0.85 &    0.11(1)  & \\
B0950+08   & J0953+0755   & 3.952 &   $-$3.59 &   0.84 &    0.07(3)  & \\
B1714$-$34 & J1717$-$3425 & 1.524 &  $-$22.75 &   0.79 &    0.26(4)  & \\
B1907+00   & J1909+0007   & 0.983 &   $-$5.33 &   0.75 &    0.15(2)  & \\
B1828$-$11 & J1830$-$1059 & 2.469 & $-$365.68 &   0.71 &    0.73(2)$^{b}$  &   100\% change in W$_{10}$ \\
B1826$-$17 & J1829$-$1751 & 3.256 &  $-$58.85 &   0.68 &    0.33(2)  & \\
B0919+06   & J0922+0638   & 2.322 &  $-$73.96 &   0.68 &    0.62(4)  & \\
B0740$-$28 & J0742$-$2822 & 5.996 & $-$604.36 &   0.66 &    2.70(20) &   20\% change in W$_{75}$ \\
B1929+20   & J1932+2020   & 3.728 &  $-$58.64 &   0.31 &    0.59(2)  & \\
\hline
\end{tabular}
\label{ta:psr_params}
\end{center}
$^{a}$Data from reference \cite{klo+06}. \\
$^{b}$Note the presence of a second harmonic at F=1.47(2) yr$^{-1}$
seen in Fig.~7 and discussed in \cite{sls00}. \\
\end{table*}

\section*{Pulse shape variations}

Following the implications of the study of PSR~B1931+24 that changes
within the magnetosphere are responsible for variations in both the
spin-down rate and the emission process\cite{klo+06}, we have sought
changes in the pulse shapes of those pulsars which have shown the
greatest fractional changes in spin-down rate in the timing noise
study.  Six pulsars show changes in pulse shape which are clearly
visible (Fig.~3).  From inspection of the profiles in Fig.~3, for each
pulsar we selected the simplest ``shape parameter'' which would
discriminate between the two extreme pulse-shape states, such as
W$_{\rm 10}$, W$_{\rm 50}$ or W$_{\rm 75}$, the full widths at 10\%,
50\% or 75\%, or W$_{\rm eq}$, the equivalent width (see the
Supporting Online Material for details on how these were determined
and their implications for the timing residuals). For six pulsars, the
observed changes in $\dot{\nu}$ are indeed directly related to changes
in pulse shape (Fig.~4). In most cases, the two quantities clearly
track one another and there is strong evidence for either correlation
or anti-correlation in all six cases (Fig.~9). It is not clear
whether the imperfect correlations are intrinsic or arise from the
sparse sampling of the time series or measurement errors.

Some of the pulse profiles suggest that increased
$|\dot{\nu}|$ is associated with increased amplitude of the central
(often described as ``core'') emission relative to the surrounding (or
``conal'') emission (Fig.~3).  PSR~B1822$-$09 exhibits a main pulse, a
precursor and an interpulse\cite{fmw81,mgb81}.  For the high
$|\dot{\nu}|$ state the precursor is weak and the interpulse is
strong, the reverse occurring for the smaller $|\dot{\nu}|$ state.
Clearly, some changes in $\dot{\nu}$ are associated with large profile
changes (e.g. PSRs~J2043+2740 and B1822$-$09) while smaller
profile changes are also observable if sufficiently high-quality
data are available (e.g. PSR~B1540$-$06).

While the main impression given by the traces in Fig.~4 is that they
are bounded by two extreme levels, there are substantial, and often
repeated, subtle changes which are synchronised in both shape
parameter and $\dot{\nu}$.  The shape parameters for the observations
of PSRs B1822$-$09, B1828$-$11 and B2035+36 imply that they spend most
of the time in just one extreme state or the other.  This is
essentially the phenomenon of mode-changing, which has been known
since shortly after the discovery of
pulsars\cite{bac70a,lyn71a,msfb80,fmw81}. In those papers, pulsars are
reported to show stable profiles, but suddenly switch to another
stable mode for times ranging from minutes to hours.  However, the
time-averaged values of the shape parameters depend upon the mix of
the two states over the averaging period and that varies with time,
causing the slower changes in the shape-parameter curves and the
spin-down rate curves.  $\sim$2500~d of detail in the shape parameters
and spin-down rates of PSRs~B1822$-$09 and B1828$-$11 (Fig.~5)
illustrate how a slowly-changing mix of the two states is reflected in
the form of the smoothed shape curves.  In PSR B1822$-$09, the events
centered on MJDs 51100 and 52050 are the sites of slow
glitches\cite{zww+04,sha07} which we confirm are not a unique
phenomenon\cite{hlk10}, but arise from short periods of time spent
predominantly in a small-$|\dot{\nu}|$, large-precursor mode.

\section*{Discussion}

The large number of pulsars observed over many years in the Jodrell
Bank data archive has allowed the identification of a substantial
number of pulsars that have large $\dot{\nu}$ changes, some of which
also have detectable, correlated pulse-shape changes.  This correlation
indicates that the causes of these
phenomena are linked and are magnetospheric in origin.  The physical
mechanism for this link is likely to be that suggested to explain the
relationship between spin-down rate and radio emission in B1931+24,
namely a change in magnetospheric particle current flow\cite{klo+06}.
An enhanced flow of charged particles causes an increase in the
braking torque on the neutron star and also in the emission radio
waves.

The link between the spin-down rate and radio-emission properties has
not been established previously, mainly because the timescales of the
long-established phenomena of mode-changing and pulse-nulling were
much shorter than the time required to measure any change in
$\dot{\nu}$.  The extended high-quality monitoring of many pulsars has
now revealed long-term manifestations of these phenomena and allowed
their unambiguous association with the spin-down rates of pulsars,
seen as timing noise.  Pulsars can spend long periods of time in one
magnetospheric state or another or in some cases switch rapidly back
and forth between states, the fractions of time spent in the two
states often varying with time.  It has long been suspected that
mode-changing and nulling are closely related
[e.g. \cite{ls05a,wmj07}].  The intermittent pulsar B1931+24 has the
largest fractional change in $\dot{\nu}$ in Table 1 and, as it
completely disappears, also has the largest apparent change in pulse
shape. Mode-changing and nulling therefore probably differ only in the
magnitude of the changes in the magnetospheric current flows.  There
is a close linear relationship between $\Delta\dot{\nu}$ and the
spin-down rate $|\dot{\nu}|$ (Fig.~10), indicating that the value of
$\dot{\nu}$ switches by about 1\% of the mean value, independent of
its magnitude.

We must also emphasise that: (1) the fast change between the states
indicates that the magnetospheric state changes on a fast time scale,
but can then be stable for many months or years before undergoing
another fast change, (2) whatever the cause of the state-switching,
for most pulsars, it is not driven by a highly periodic (high-Q)
oscillation and (3) increased $|\dot{\nu}|$ is associated with
increased amplitude of the core emission relative to conal emission.
The fast state-changes seem to rule out free precession as the origin
of the oscillatory behaviour.  PSR~B1828$-$11 was considered unique in
that it was the only pulsar that showed clear evidence for free
precession \cite{sls00}.  Our model indicates that this pulsar is not
unique and exhibits the same state-changing phenomenon shown here for
other pulsars.

If we could monitor a pulsar continuously, then its magnetospheric
state at any given time could be determined from the pulse shape. The
state gives a measure of the spin-down rate, allowing the timing noise
to be removed (an example is given in Fig.~6). The most stable
millisecond pulsars are being regularly observed from many
observatories world-wide in the hope of making the first direct
detection of gravitational waves \nocite{hob08}.  The first-discovered
millisecond pulsar, PSR B1937$+$21, can be timed with high precision
(of $\sim 100$\,ns) over short data spans, but low-frequency timing
irregularities dominate the timing residuals over data spanning more
than $\sim 3$\,yr \cite{ktr94} making this pulsar potentially unusable
for gravitational wave detection experiments.  However, if
magnetospheric state switching is responsible and can be applied to
millisecond pulsars, then the timing irregularities can be modelled
and removed, raising the possibility of producing an essentially
stable clock.



\begin{scilastnote}
\item Pulsar research at JBCA is supported by a Rolling Grant from the UK
Science and Technology Facilities Council.\\
GH is the recipient of an Australian Research Council QEII Fellowship (\#DP0878388).\\
MK is supported by a salary from the Max-Planck Society.\\
Pulsar research at UBC is supported by an NSERC Discovery Grant.\\
\end{scilastnote}


\clearpage

\noindent {\bf Fig. 1.}  Pulsar timing residuals relative to a simple
spin-down model of the pulse frequency and its first derivative.  For
PSRs~B0919$-$06, B1540$-$06 and B1828$-$11 we have also included the
frequency second derivative in the model. For each pulsar, the
peak-to-peak range in residual is given, and the vertical scale has
been adjusted to give the same peak-to-peak deflection in the
diagram. We use data updated from those presented in \cite{hlk10} and also
include data for PSR~J2043+2740. The residuals were obtained using the
\textsc{tempo2} software package\cite{hem06}.

\noindent {\bf Fig. 2.}  Variations in the spin-down rate $\dot{\nu}$
for 17 pulsars during the past 20 years.  We determined these values
by selecting small sections of data of length $T$ and fitting for
values of $\nu$ and $\dot{\nu}$, repeating at intervals of $\sim$T/4
through each data set. The chosen value of $T$ is the smallest
required to provide sufficient precision in $\dot{\nu}$ and is given
below each pulsar name. $T$ is typically 100-400\,d so that any
short-time-scale variations will be smoothed out.  For each pulsar, we
adjusted the vertical scale to give the same peak-to-peak amplitude
and subtracted an arbitrary vertical offset.  Because $\dot{\nu}$ is
negative, an increase in the rate of spin-down is represented by a
downward deflection in this diagram.

\noindent {\bf Fig. 3.}
The integrated profiles at 1400~MHz of six pulsars which show
long-term pulse-shape changes.  For each pulsar, the two
traces represent examples of the most extreme pulse shapes observed.
The profile drawn in the thick line corresponds to the
largest rate of spin-down $|\dot{\nu}|$.  The profiles are scaled so
that the peak flux density is approximately the same. PSR~B1822$-$09
has an interpulse which is displayed, shifted by half the pulse period,
in the second trace below the main pulse.

\noindent {\bf Fig. 4.}
The average value of pulse shape parameter and spin-down
rate $\dot{\nu}$ measured for six pulsars.  The lower trace in each
panel (right-hand scale) shows the same values of $\dot{\nu}$ given in
Fig.~2, while the upper trace gives a measure of the
pulse shape, with the scale given to the left.  W$_{10}$, W$_{50}$ and
W$_{75}$ are the full widths of the pulse profile at 10\%, 50\% and
75\% of the peak pulse amplitude respectively, W$_{eq}$ is the pulse
equivalent width, the ratio of the area under the pulse to the peak
pulse amplitude, and A$_{pc}$/A$_{mp}$ is the ratio of the amplitudes
of the precursor and main pulse. The time over which a shape parameter
is averaged is the same as the time $T$ given in
Fig.~2 for the fitting of $\dot{\nu}$. The
uncertainty on a shape parameter is derived from the standard
deviation of the individual values used to determine the average.

\noindent {\bf Fig. 5.}
The variations in pulse shape parameters for PSR~B1822$-$09
(a-c) and PSR~B1828$-$11 (d-f). Traces a, c, d and f are taken from
Fig.~4 and show the smoothed values of shape
parameter and spin-down rate for the two pulsars, while diagrams b and
e show the values of shape parameter for individual observations which
are typically between 6 and 18 minutes duration.  Note that for both
pulsars, individual shape parameter values typically take either a
high or low value. 

\begin{figure*}
\begin{center}
\includegraphics[width=12cm]{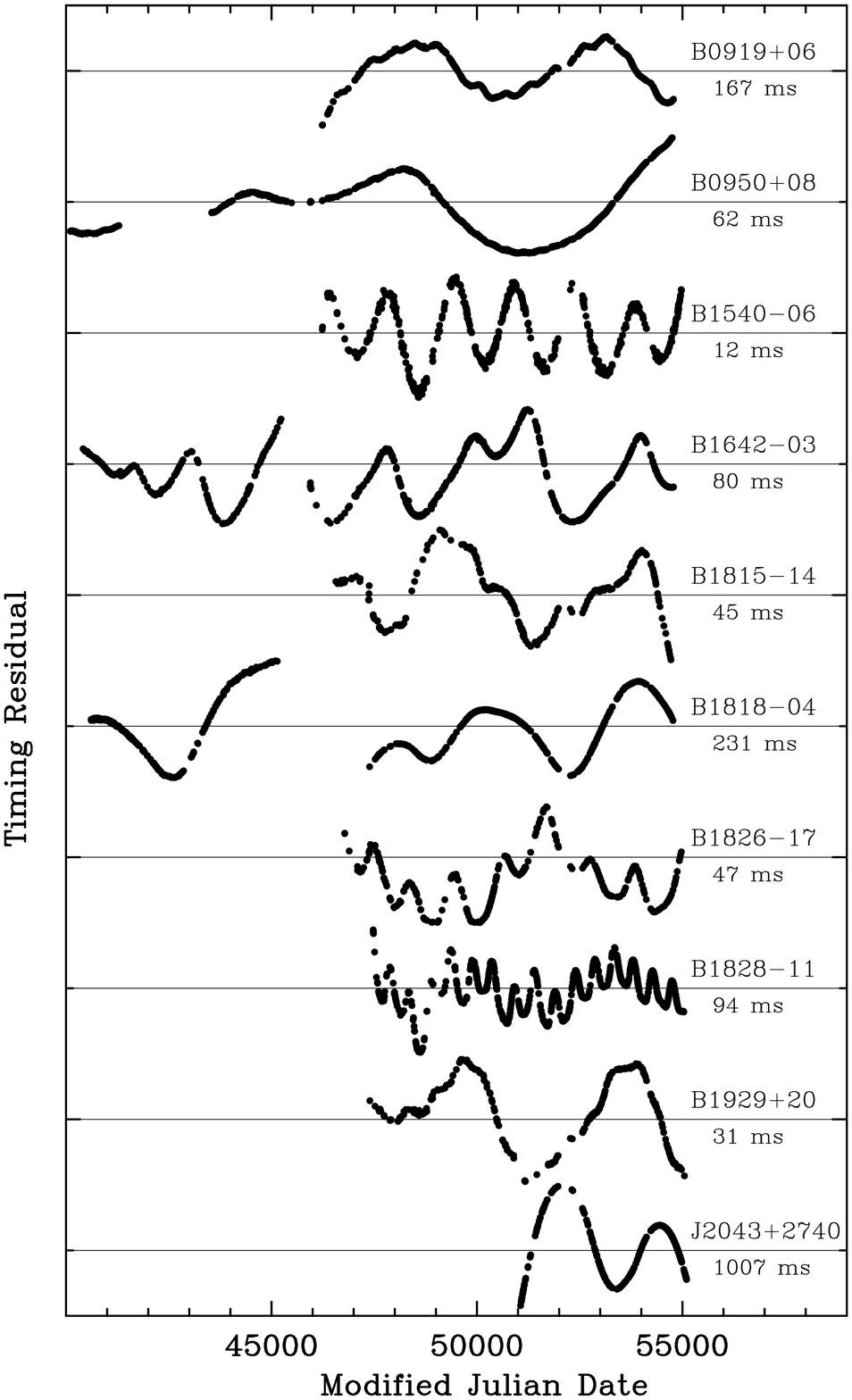}
\end{center}
\caption[]{}
\end{figure*}

\begin{figure*}
\begin{center}
\includegraphics[width=12cm]{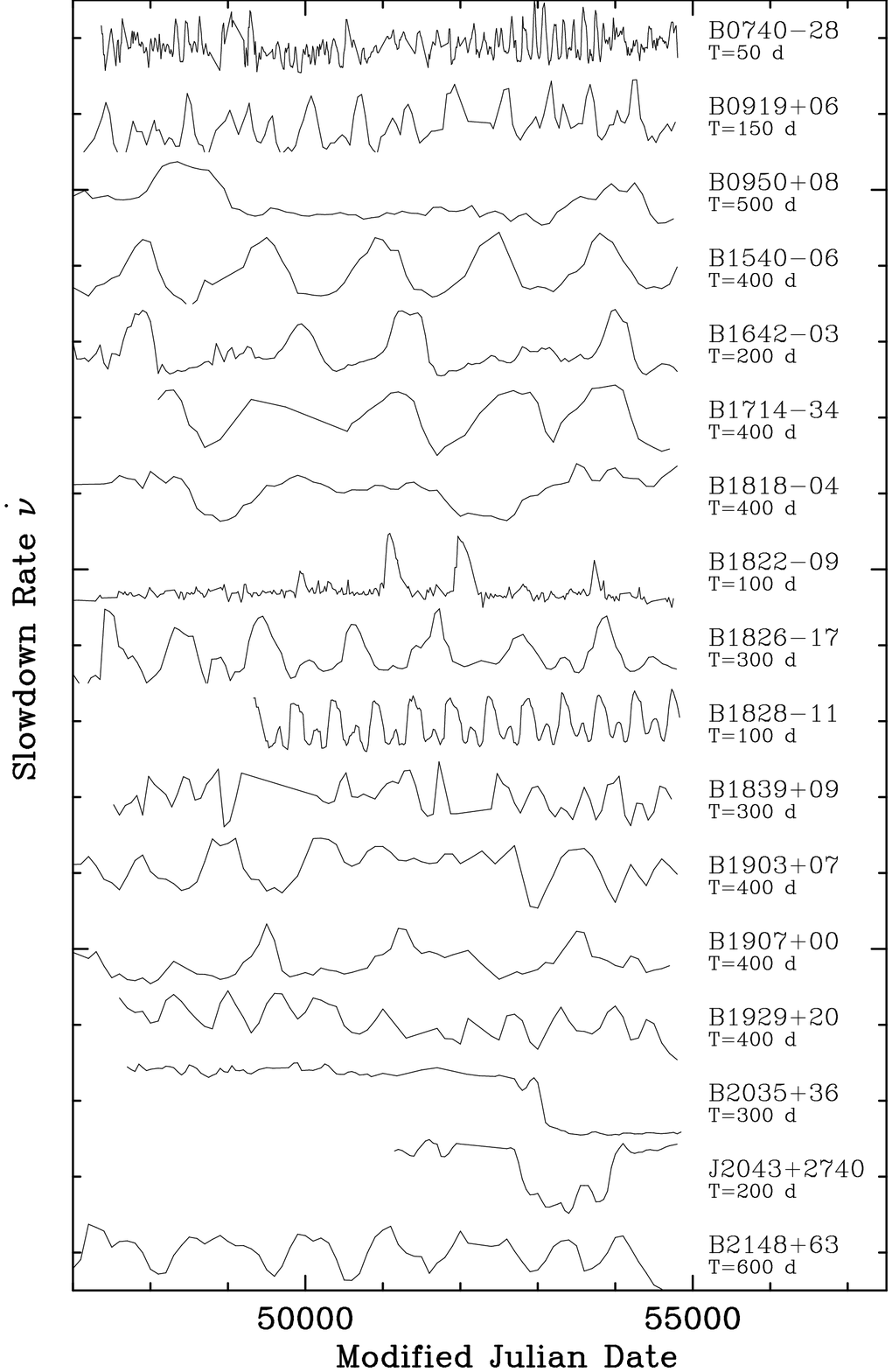}
\end{center}
\caption[]{}
\end{figure*}

\begin{figure*}
\begin{center}
\includegraphics[width=12cm]{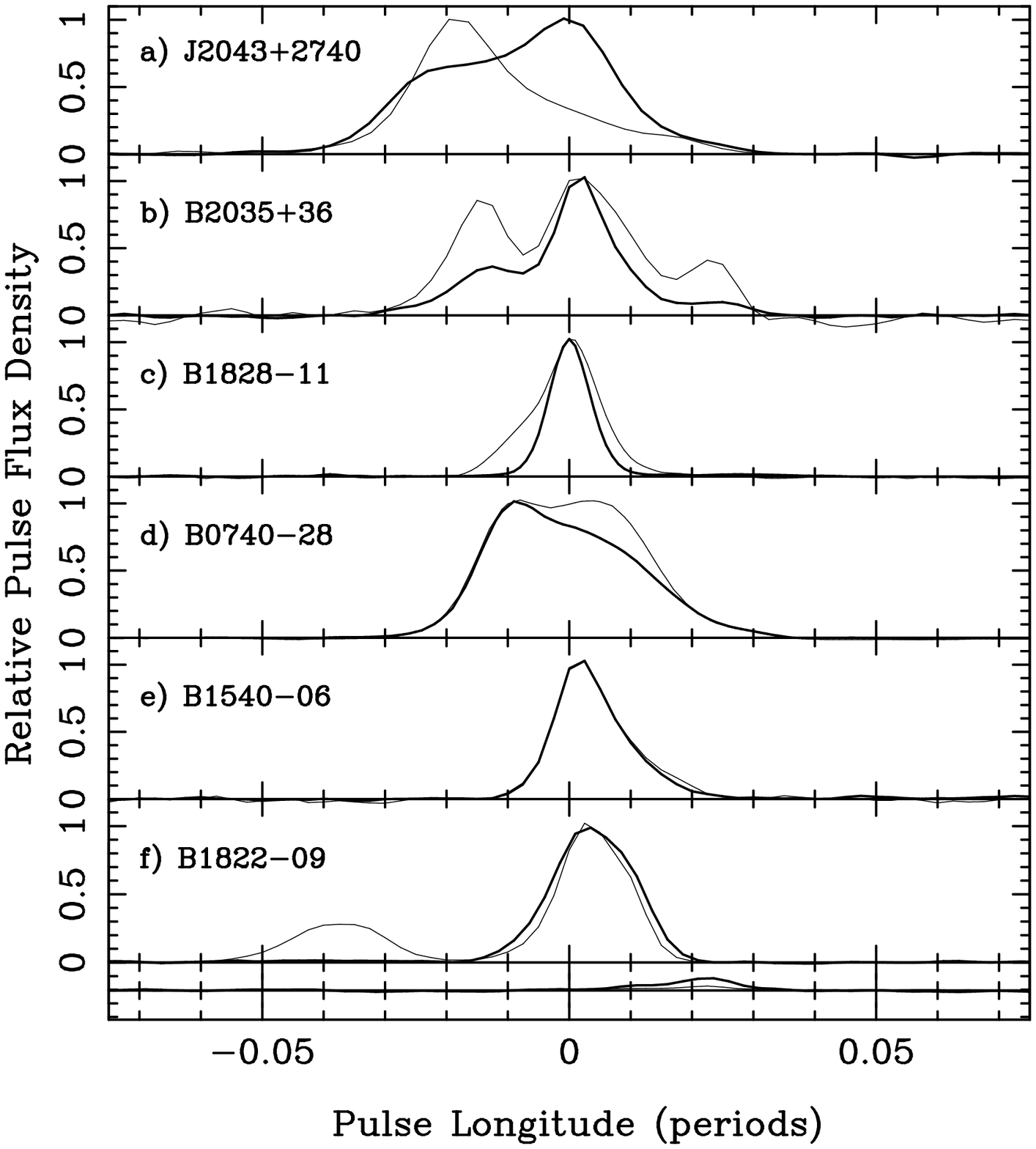}
\end{center}
\caption[]{}
\end{figure*}

\begin{figure*}
\begin{center}\includegraphics[width=12cm]{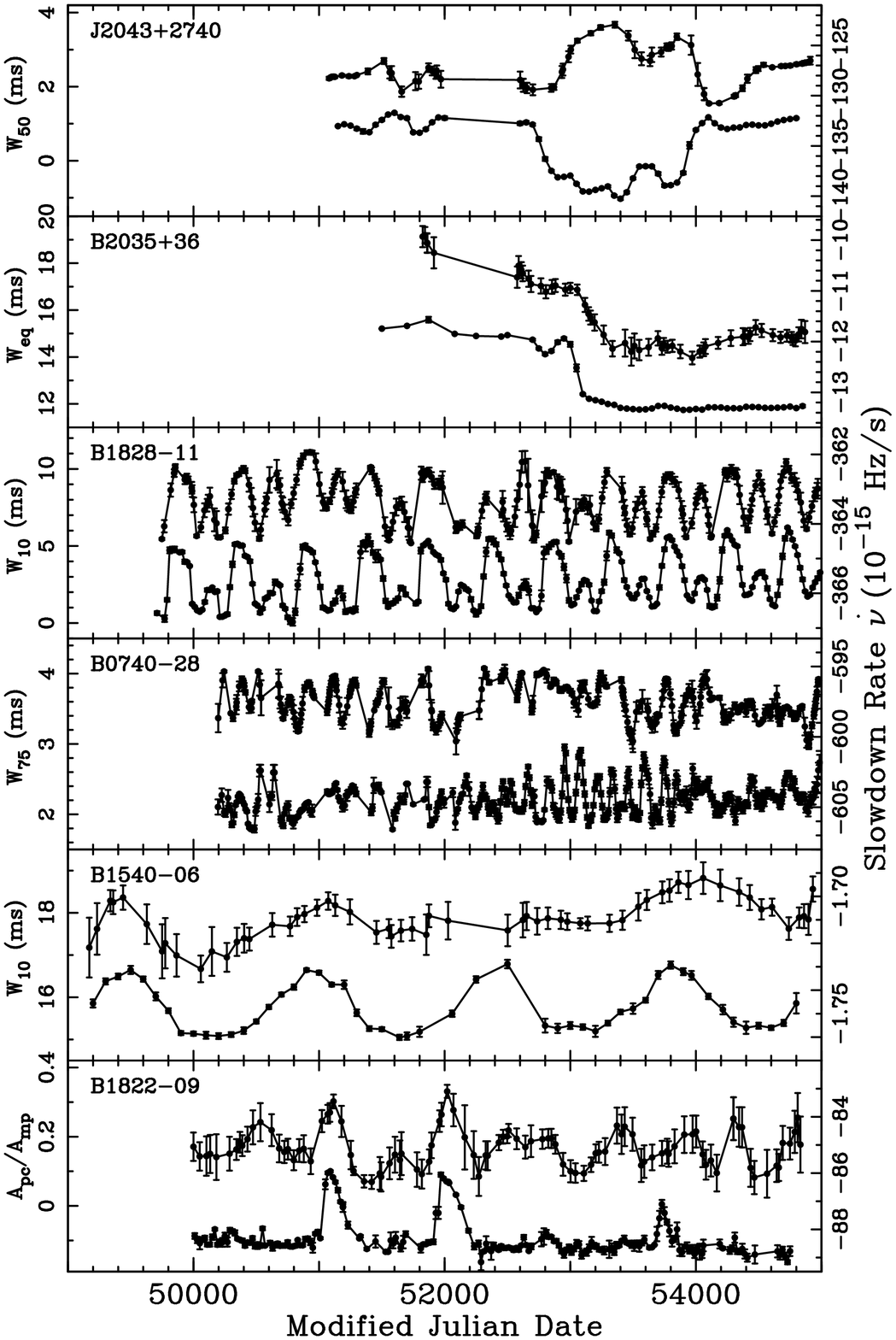}\end{center}
\caption[]{}
\end{figure*}

\begin{figure*}
\begin{center}\includegraphics[width=12cm]{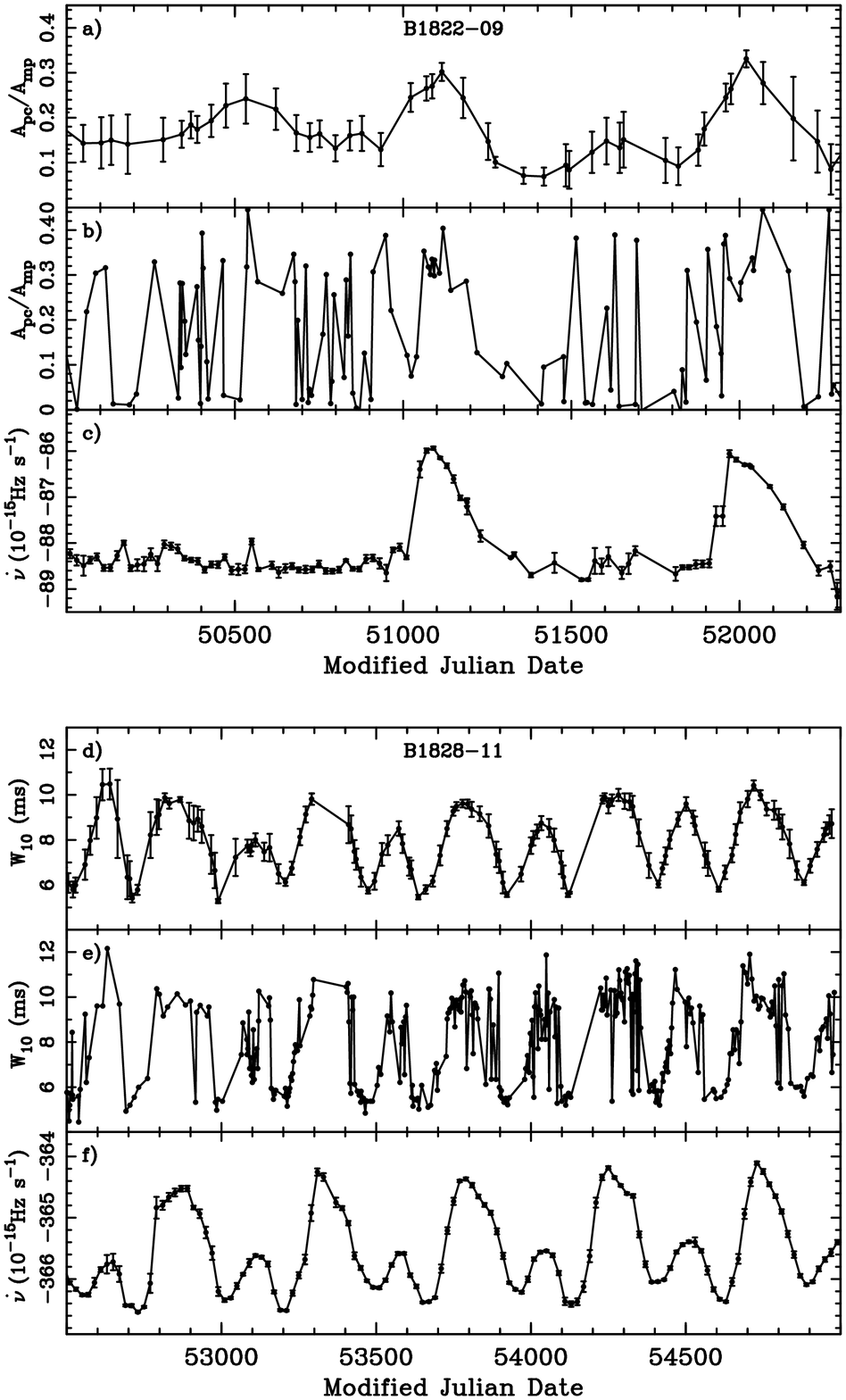}\end{center}
\caption[]{}
\end{figure*}

\clearpage



\setcounter{page}{1}

\section*{Supporting Online Material}

\subsection*{Determining pulse shape parameters}
The pairs of pulse profiles in Fig.~3 of the journal paper were inspected
and fitted with two or three gaussian components, as required to
provide satisfactory descriptions of the profiles. The same components
were fitted to each observed profile, and a synthetic profile
produced, from which the value of the chosen shape parameter used in
Fig.~4 of the journal paper was determined.  W$_{10}$, W$_{50}$ and
W$_{75}$ are the widths measured at 10\%, 50\% and 75\% of the pulse
peak. W$_{\rm eq}$ is the equivalent width, being the area under the
pulse divided by the peak amplitude, and A$_{\rm pc}$/A$_{\rm mp}$ is
the ratio of the peak amplitudes of the precursor and main pulse
components.  Note that this procedure has the virtue of applying a
quasi-optimum filter to the data in order to minimise the effects of
high-frequency noise on the values of the parameters.

\subsection*{Determining times-of-arrival}
We note that the times-of-arrival used to derive the timing residuals
of Fig.~1 in the journal paper were usually obtained by matching the
observed pulse profiles with templates derived from average profiles
obtained over large sections of the data. Hence, for the objects in
Fig.~3 of the journal paper, the templates have intermediate shapes,
and are usually not perfectly matched to the observed profile, giving
rise to possible systematic offsets in the timing residuals.  The
maximum magnitudes of the offsets have been estimated by comparing the
times-of-arrival obtained by fitting the two profiles in Fig.~3 with
the corresponding template. For the six objects, the differences in
the offsets were respectively 0.70, 0.05, 0.25, 0.17, 0.04, and 0.14
ms.  Since these are much smaller than the variations seen in Fig.~1
of the journal paper, we conclude that the profile switching and use
of a single template has little impact upon the structures seen in the
timing residuals.  In practice, for PSR~B1828$-$11 the procedure
described in ({\it Ref 1 of SoM}) was used and therefore the TOAs are not
prone to such systematic errors.

\subsection*{Observational limitations}
Within the relatively small number of the pulsars in our sample which
have high signal-to-noise ratio profiles, pulse-shape variations are
observed which are correlated with spin-down rate. The results show
that multiple pulse-profile and associated spin-down states that
switch on timescales of weeks to years is a common phenomenon seen in
many pulsars.  We believe that it is possible that all pulsars which
display timing noise may show the same correlated pulse-shape
behaviour, although it is so far unobserved in most pulsars, because
of a combination of modest profile changes, poor signal-to-noise ratio
and often relatively-short available data spans. The profile changes
may be small, for instance, if changes in the pulsar emission beam
happen to be small in that part of the beam which crosses the
line-of-sight to the Earth.

\subsection*{Simulations of timing residuals (Fig.~6)}
We have calculated some simulations of a two-state spin-down model for
pulsar timing noise in which only two parameters determine the form of
the simulations, namely the ratio R of time spent in high and low
spin-down states, and the rms fractional dither D in the switching
period. In detail, simulated values of $\dot{\nu}$ were determined for
a regular time sampling. $\dot{\nu}$ switched between two modes
$\dot{\nu}_{\rm 1}$ and $\dot{\nu}_{\rm 2}$, spending a time t$_{\rm
1}$ in the first mode and t$_{\rm 2}$ in the second mode (typically
t$_{\rm 1}$ and t$_{\rm 2}$ are a few hundred days).  The ratio of
t$_{\rm 1}$ and t$_{\rm 2}$ equals R.  For simulations involving
dithering the timescales, t$_{\rm 1}$ and t$_{\rm 2}$ are slightly
modified by adding a Gaussian random deviate with an rms of Dt$_{\rm
1}$ and Dt$_{\rm 2}$ respectively.  The resulting $\dot{\nu}$ values
are numerically integrated twice to produce pulse phase, followed by
appropriate sampling and a quadratic polynomial removed to form the
resulting simulated timing residuals.  Note how dither in the
switching period can give rise to low-frequency structure in the
residuals (Fig.~6c).

\begin{figure*}
\begin{center}\includegraphics[width=16cm,angle=0.0]{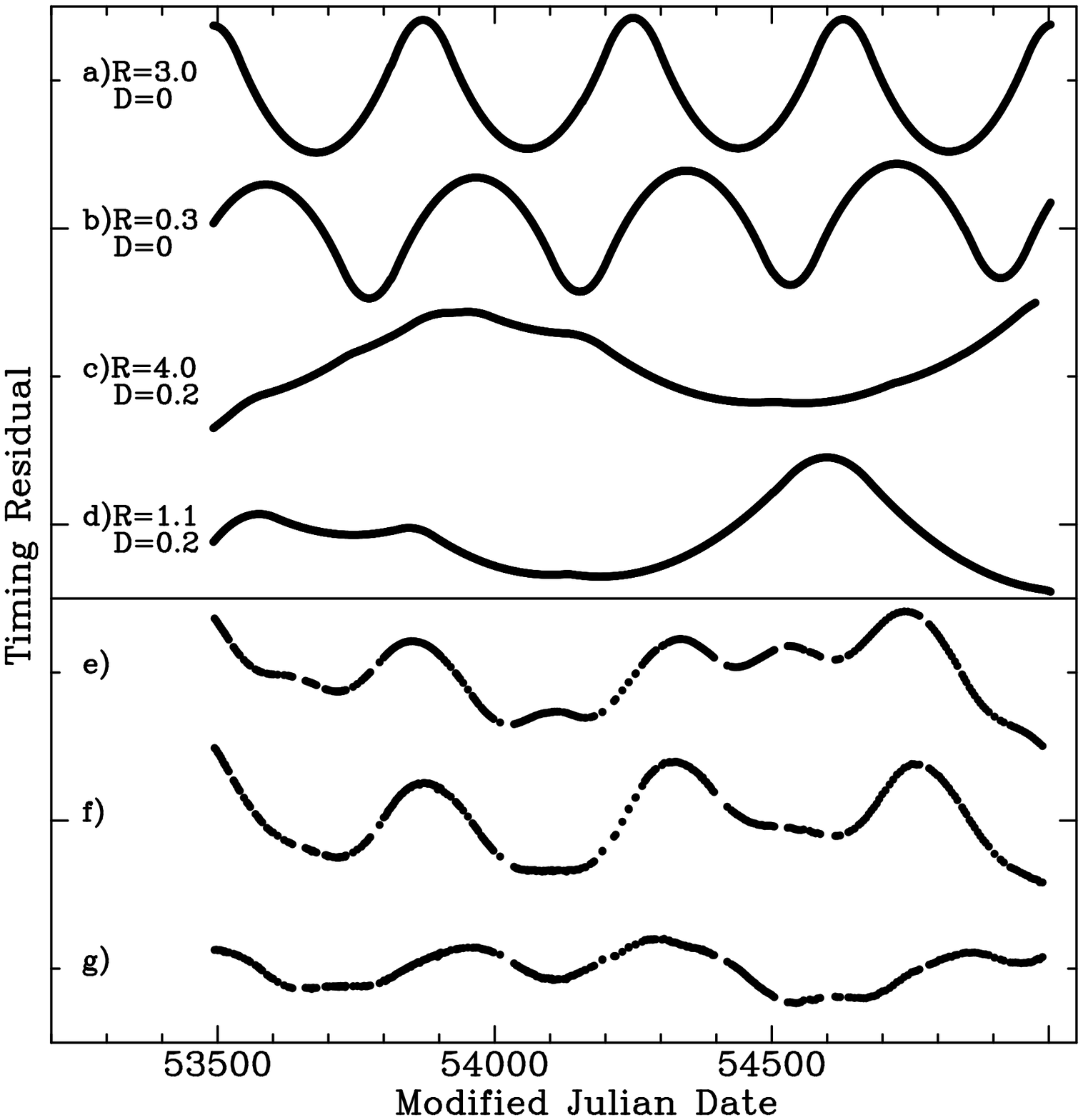}\end{center}
\caption{ Simulations of timing residuals. a)-d) Simulations of a
two-state spin-down model for pulsar timing noise in which only two
parameters determine the form of the simulations, namely the ratio R
of time spent in high and low spin-down states, and the rms fractional
dither D in the switching period (See supporting text). Note how
dither in the switching period can give rise to low-frequency
structure in the residuals.  e) simulated timing residuals for
PSR~B1828$-$11 in which the spin-down state is determined purely from
the observed pulse shape parameter f) observed timing residuals
for PSR~B1828$-$11 from a simple spin-down model, which shows most of
the features predicted by e), and g) the difference between the
observed and simulated timing residuals. In spite of the severe
undersampling of the shape parameter due to telescope availability
($<1\%$ of the time), this demonstrates how it might be possible
to ``correct'' the times-of-arrival for spin-state variations
indicated by the pulse shape.}
\label{fg:simulations}
\end{figure*}

\begin{figure*}
\begin{center}\includegraphics[width=12cm]{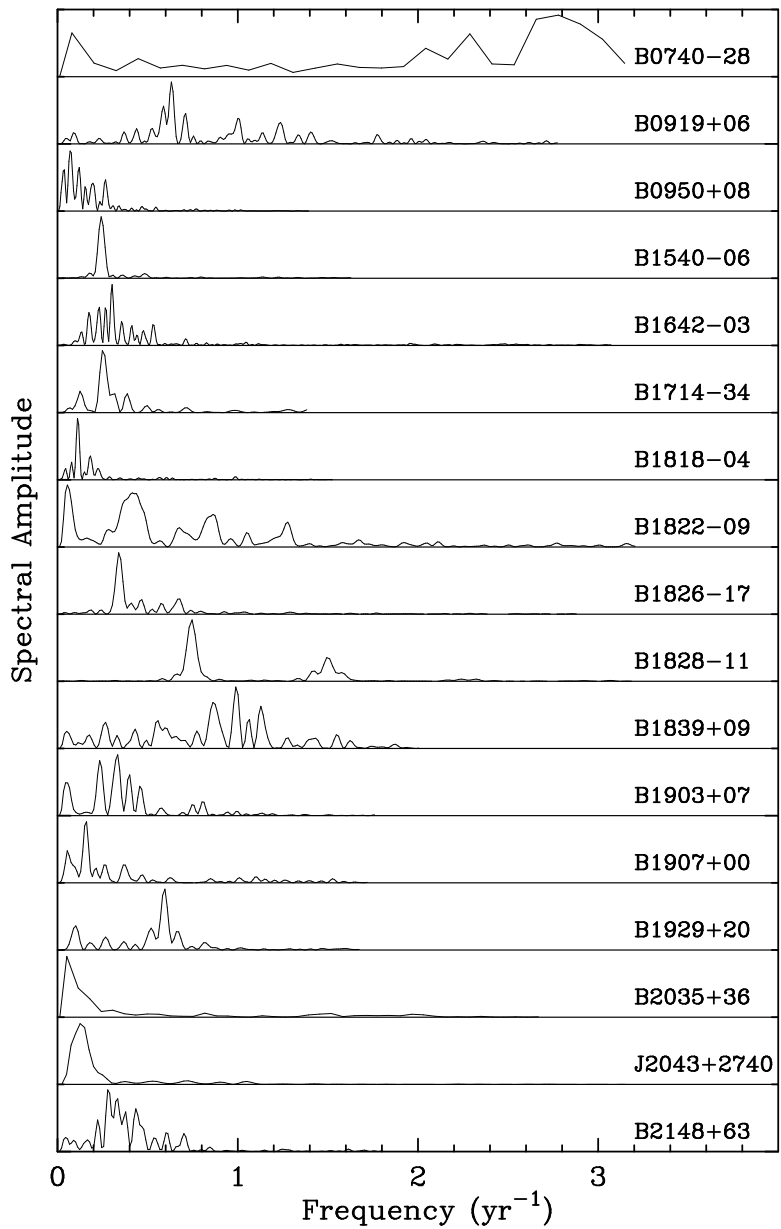}\end{center}
\caption{ Lomb-Scargle spectra of the
spin-down rates $\dot{\nu}$ presented in
Fig.~2 of the journal paper for 17 pulsars. The peaks of
all the spectra have been normalised to the same amplitude. }
\label{fg:f1_ls}
\end{figure*}

\begin{figure*}
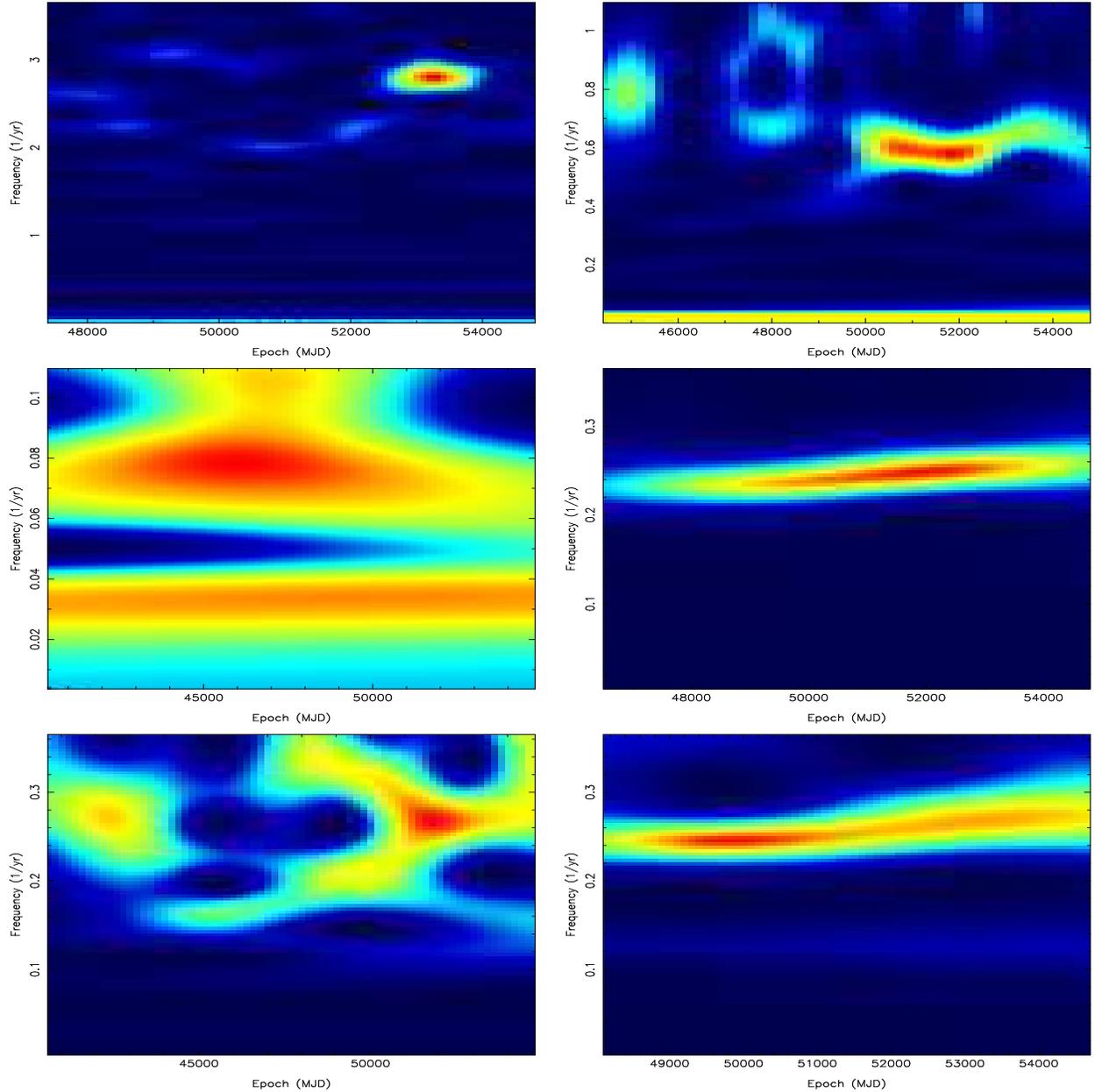

\begin{center}
\begin{tabular}{ccccc} 
\mbox{\psfig{file=0740.ps,height=54mm,width=80mm,angle=-90}} &
\mbox{\psfig{file=0919.ps,height=54mm,width=80mm,angle=-90}} \\
\mbox{\psfig{file=0950.ps,height=54mm,width=80mm,angle=-90}} &
\mbox{\psfig{file=1540.ps,height=54mm,width=80mm,angle=-90}} \\
\mbox{\psfig{file=1642.ps,height=54mm,width=80mm,angle=-90}} &
\mbox{\psfig{file=1714.ps,height=54mm,width=80mm,angle=-90}} \\  
\end{tabular}
\end{center}
\caption{a) Wavelet spectra for pulsars B0740$-$28 and B0919+06 (top
row), B0950+08 and B1540$-$06 (middle row) and B1642$-$03 and
B1714$-$34 (bottom row).  The frequency ranges shown cover the
periodicities suggested in Fig.~7. The wavelet Z-statistic is computed
as a function of both time and frequency.  The wavelet "window" can be
specified by a "decay constant", $c$, that defines the number of
cycles of a given frequency $f$ expected within the window. Values
between 0.001 and 0.01 were chosen in an attempt to obtain the best
compromise between frequency and time resolution, given the data in
Fig.~2.  The results agree well with the periodicities derived from
the Lomb-Scargle analysis but demonstrate that for some sources the
effective frequencies are varying or not always present.}
\end{figure*}\addtocounter{figure}{-1}

\begin{figure*}
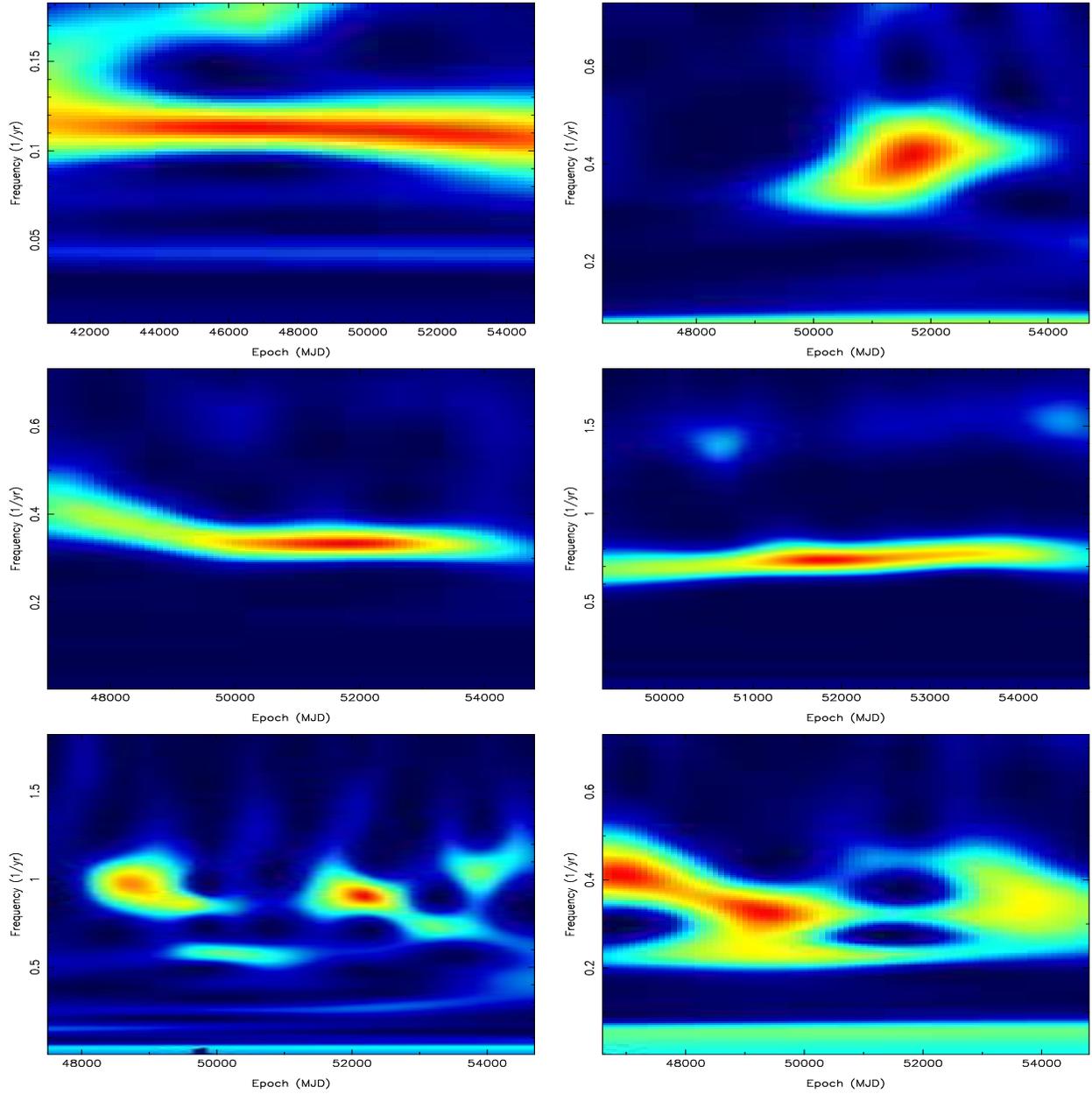

\begin{center}
\begin{tabular}{ccccc} 
\mbox{\psfig{file=1818.ps,height=54mm,width=80mm,angle=-90}} &  
\mbox{\psfig{file=1822.ps,height=54mm,width=80mm,angle=-90}} \\ 
\mbox{\psfig{file=1826.ps,height=54mm,width=80mm,angle=-90}} &
\mbox{\psfig{file=1828.ps,height=54mm,width=80mm,angle=-90}} \\
\mbox{\psfig{file=1839.ps,height=54mm,width=80mm,angle=-90}} &
\mbox{\psfig{file=1903.ps,height=54mm,width=80mm,angle=-90}} \\
\end{tabular}
\end{center}
\caption{b) Wavelet spectra for pulsars B1818$-$04 and B1822$-$09 (top
row), B1826$-$17 and B1828$-$11 (middle row) and B1839+09 and
B1903+07 (bottom row).}
\end{figure*}\addtocounter{figure}{-1}

\begin{figure*}
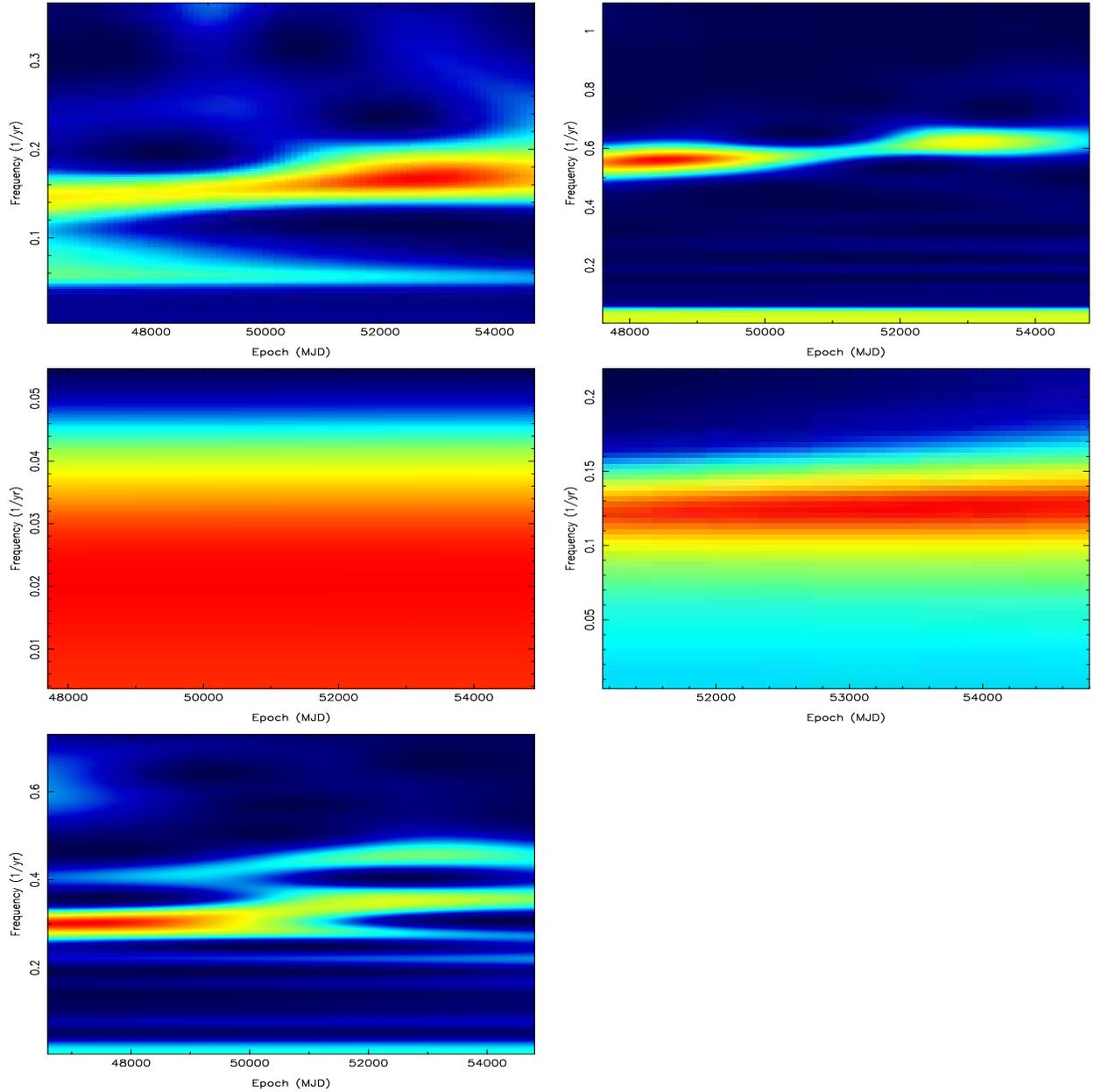

\begin{center}
\begin{tabular}{ccccc} 
\mbox{\psfig{file=1907.ps,height=54mm,width=80mm,angle=-90}} &
\mbox{\psfig{file=1929.ps,height=54mm,width=80mm,angle=-90}} \\
\mbox{\psfig{file=2035.ps,height=54mm,width=80mm,angle=-90}} &
\mbox{\psfig{file=2043.ps,height=54mm,width=80mm,angle=-90}} \\
\mbox{\psfig{file=2148.ps,height=54mm,width=80mm,angle=-90}} \\
\end{tabular}
\end{center}
\caption{c) Wavelet spectra for pulsars B1907+00 and B1929+20 (top
row), B2035+36 and J2043+2740 (middle row) and B2148+63 (bottom row).}
\end{figure*}

\begin{figure*}
\begin{center}\includegraphics[width=12cm]{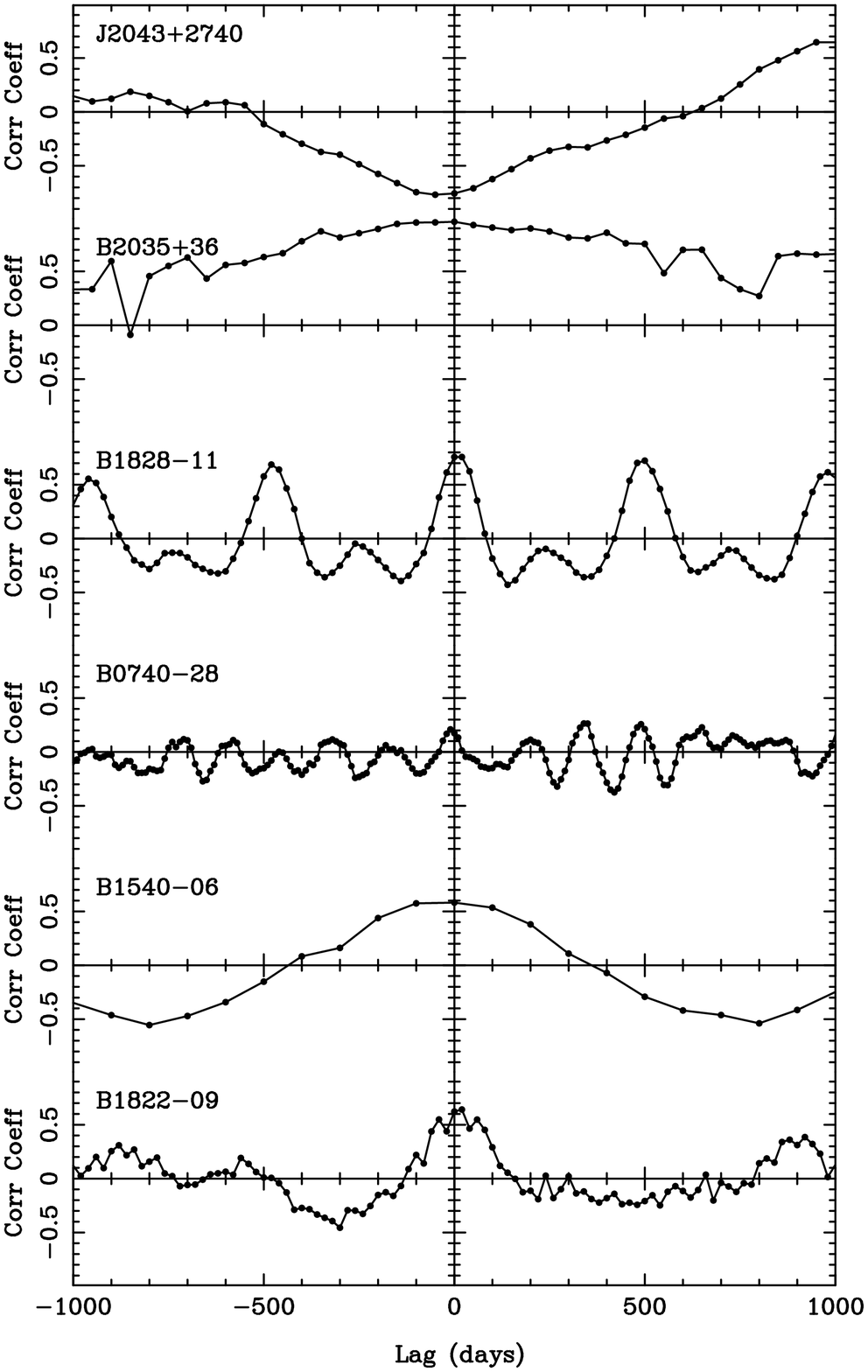}\end{center}
\caption{ Cross-correlation functions between the average values of
pulse shape parameters and the spin-down rates $\dot{\nu}$ shown in
Fig.~4 of the journal paper for six pulsars. Note that in all cases,
the magnitude of correlation coefficient always peaks close to
zero lag. }
\label{fg:f1_shape_cc}
\end{figure*}

\begin{figure*}
\begin{center}\includegraphics[width=12cm]{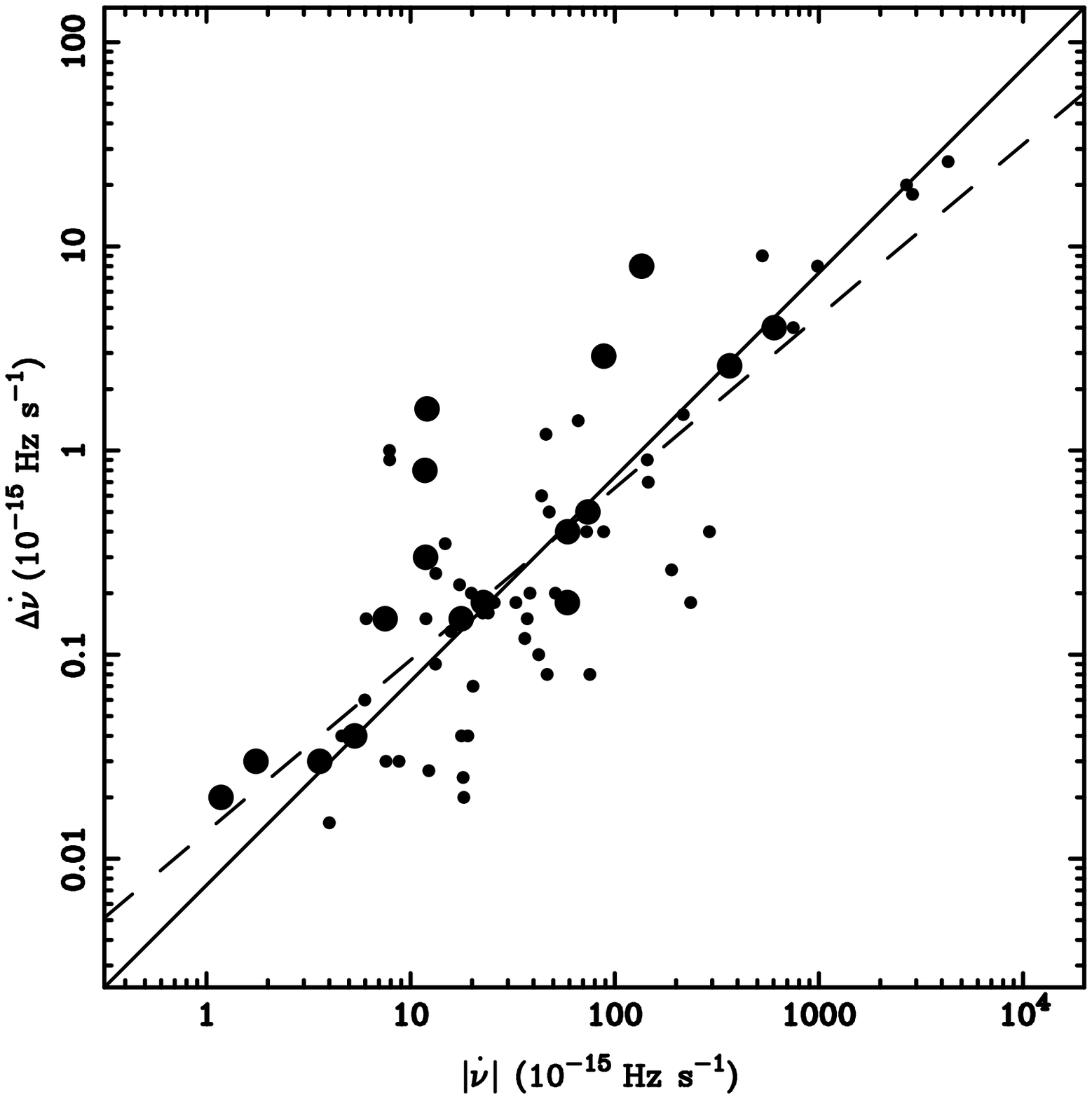}\end{center}
\caption[]{ The dependence of the magnitude of the switching in
$\dot{\nu}$ upon the magnitude of the average spin-down rate
$|\dot{\nu}|$ for the 17 pulsars in Table~1 excepting PSR~B1931+24
(large symbols) and another 51 pulsars which were studied in ({\it Ref
2 of SoM}) (small
symbols). There is an excellent correlation between the logarithm of
the variables with a correlation coefficient of 0.80. The solid line
has a slope of unity and the dashed line is the best-fitting straight
line, which has a slope of 0.84$\pm$0.08. }
\label{fg:df1_f1}
\end{figure*}

\end{document}